\documentclass[12pt]{article}
\usepackage{amstex}
\usepackage{amssymb}
\usepackage{babel}
\usepackage{latexsym}
\begin{document}
\newfont{\fraktgm}{eurm10 scaled 1728}
\newcommand{\graktur}{\baselineskip12.5pt \fraktgm}
\newfont{\fraktfm}{eurm10 scaled 1440}
\newcommand{\frakture}{\baselineskip12.5pt\fraktfm}
\newfont{\fraktrm}{eurm10}
\newcommand{\fraktur}{\baselineskip12.5pt\fraktrm}
\newfont{\fraktem}{eurm6}
\newcommand{\fraktr}{\baselineskip12.5pt\fraktem}
\protect\newtheorem{principle}{Principle}[section]
\protect\newtheorem{theo}[principle]{Theorem}
\protect\newtheorem{prop}[principle]{Proposition}
\protect\newtheorem{lem}[principle]{Lemma}
\protect\newtheorem{co}[principle]{Corollary}
\protect\newtheorem{de}[principle]{Definition}
\newtheorem{ex}[principle]{Example}
\begin{titlepage} \enlargethispage{4cm} 
\vspace*{-1.8cm}
\vskip.6in 
\begin{center} 
{\graktur On Tracial Operator Representations of Quantum 
\\ \vskip.06in} 
{\graktur Decoherence Functionals} 
\vskip.75in 
{{\Large \frakture Oliver Rudolph} $^a$} 
\vskip.3in 
{\normalsize \sf II.~Institut f\"ur Theoretische Physik}
\vskip.05in
{\normalsize \sf Universit\"at Hamburg} 
\vskip.05in 
{\normalsize \sf Luruper Chaussee 149} 
\vskip.05in 
{\normalsize \sf D-22761 Hamburg, Germany}
\vskip.3in
{\normalsize \sf and}
\vskip.3in
{{\Large {\frakture J}.~{\frakture D}.~{\frakture 
Maitland Wright}} $^b$} 
\vskip.3in 
{\normalsize \sf Analysis and Combinatorics Research 
Centre}
\vskip.05in
{\normalsize \sf Mathematics Department}
\vskip.05in
{\normalsize \sf University of Reading}
\vskip.05in
{\normalsize \sf Reading RG6 6AX, England}
\vskip.75in
\end{center}
\normalsize
\begin{center}
{ABSTRACT}
\end{center}
\smallskip
\noindent A general \emph{quantum history theory} can be 
characterised by the 
space of histories and by the space of decoherence 
functionals. 
In this note we consider the situation where the space 
of histories 
is given by the lattice of projection operators on an 
infinite 
dimensional Hilbert space $\mathfrak{H}$. 
We study operator representations 
for decoherence functionals on this space of histories. 
We first 
give necessary and sufficient conditions for a 
decoherence functional 
being representable by a trace class operator on 
$\mathfrak{H}
\otimes 
\mathfrak{H}$, an infinite 
dimensional analogue of the Isham-Linden-Schreckenberg 
representation 
for finite dimensions.  Since this excludes many 
decoherence 
functionals of physical interest, we then identify the 
large 
and physically important class of decoherence 
functionals which 
can be represented, canonically, by bounded operators 
on 
$\mathfrak{H}
\otimes \mathfrak{H}$. \\ 
\\ \bigskip \noindent
\centerline{\vrule height0.25pt depth0.25pt width4cm 
\hfill}
\noindent
{\footnotesize $^a$ email: rudolph@@x4u2.desy.de \\
$^b$ email: J.D.M.Wright@@reading.ac.uk} 
\end{titlepage}
\newpage
\section{Introduction}
The consistent histories approach to quantum 
mechanics has attracted much interest in recent years. 
The 
consistent or decoherent histories approach to quantum 
theory 
is a fresh, novel attempt to formulate a substantial 
generalisation 
of standard quantum mechanics which, as far as the 
mathematical 
machinery is concerned, stays remarkably close to 
ordinary Hilbert 
space quantum mechanics. It introduces new concepts 
into quantum 
mechanics and is structurally different from all 
other approaches 
to quantum mechanics. It has enriched and deepened our 
understanding 
of non-relativistic quantum mechanics quite 
generally and in 
particular of the interpretation of standard 
Hilbert space quantum 
mechanics. \\ 
There is also the hope that physical general 
quantum history 
theories may be constructed in terms of the 
concepts of the histories 
approach, which generalise standard Hilbert 
space quantum mechanics. \\ 

The consistent histories approach to non-relativistic 
quantum 
mechanics was inaugurated in a seminal paper by 
Griffiths 
\cite{Griffiths84}. In this paper Griffiths 
introduced histories 
mainly 
as a tool for formulating a realistic interpretation 
of non-relativistic 
quantum mechanics. This so-called 
`\emph{consistent histories interpretation}' 
has been further developed and brought to its present 
form 
by Omn\`{e}s \cite{Omnes88} - \cite{Omnes94}. \\ 
(In the consistent histories interpretation it is 
asserted that 
quantum mechanics provides a realistic description 
of individual 
quantum mechanical systems, regardless of whether 
they are open 
or closed and regardless of whether there is an 
external observer 
or not. Probabilities are associated with 
\emph{complete} histories 
and are thought of as measures of \emph{propensities} 
or tendencies 
inherent in the quantum system in question. The 
assignment of 
probabilities to histories in a certain set is only 
admissible 
when this set of histories carries the structure of a 
Boolean 
lattice and satisfies an additional so-called 
consistency condition. \\ 
In a series of publications Gell-Mann and Hartle 
\cite{GellMannH90a} - \cite{GellMannH93} have 
studied quantum cosmology and the path 
integral formulation of 
relativistic quantum field theory in terms of the 
concepts of 
the histories approach. \\ 
However, all the above mentioned authors 
have stuck to the 
usual Hilbert space formalism (or to the 
usual path integral 
formulation) of Hamiltonian quantum mechanics 
or Hamiltonian quantum field theory and have in 
essence considered 
only histories which are time-sequences of 
single-time events (or -- in
the path integral formulation -- classes of 
Feynman paths). \\ 
Further important developments and results can 
be found in the 
work of Dowker and Kent \cite{DowkerK96,DowkerK95}. 
Specifically, Dowker and Kent 
showed that certain hopes of the original 
Gell-Mann--Hartle 
programme cannot be fulfilled and that 
some incidental claims 
to be found in the early literature of 
the consistent histories 
approach cannot be upheld. \\ 
The investigation here is not affected by the 
negative results due to Dowker and Kent. \\ 

In an ingenious work Isham \cite{Isham94} has 
broadened both the scope 
and the mathematical framework of the consistent 
histories approach 
to quantum mechanics. Isham has formulated a 
natural algebraic 
generalisation of the consistent histories 
approach. \\ 
In his approach a general quantum history 
theory is characterised 
by the space of histories on the one hand 
and by the space of 
decoherence functionals on the other hand. 
In Isham's approach 
histories are identified with the general 
temporal properties 
of the quantum system or in a somewhat 
different language with 
the \emph{temporal events}. In general 
these temporal events or 
histories are more general objects than 
simply time-sequences 
of single-time events but must be regarded 
as events intrinsically 
spread out in time. Histories are regarded 
as fundamental entities 
in their own right. \\ 
Isham's approach has subsequently become 
the subject of intense 
study. The reader is referred to the 
original articles by Isham 
and Linden \cite{IshamL94,IshamL95,IshamL96}, 
by Isham, Linden and Schreckenberg 
\cite{IshamLS94}, 
by Schreckenberg 
\cite{Schreckenberg97,Schreckenberg96,Schreckenberg95}, 
by Isham \cite{Isham97}, by Pulmannov\'{a} 
\cite{Pulmannova95}, 
and by the present authors 
\cite{Wright95,Wright96,Rudolph96,Rudolph97}. \\ 

Dual to the notion of history is the notion of 
decoherence functional. 
The decoherence functional determines the 
consistent sets of 
histories in the theory and the probabilities 
assigned to histories 
in the consistent sets. More specifically, a 
decoherence functional 
$d$ maps every ordered pair 
of histories $h, k$ to a complex number denoted 
by $d(h,k)$. The number $d(h,k)$ is interpreted 
in physical terms 
as a measure of the mutual interference of the 
two histories 
$h$ and $k$. A consistent set of histories 
consists of histories 
whose mutual interference is sufficiently 
small, such that the 
diagonal value $d(h,h)$ can be interpreted 
as the probability of the 
history $h$ in this consistent set. \\ 
In standard quantum mechanics  the state of 
some quantum mechanical 
system comprises all probabilistic 
predictions of quantum mechanics 
for the system in question. This idea 
of the notion of the state 
can be carried over to general quantum 
history theories: it is 
in this sense that decoherence functionals 
can be said to represent 
the \emph{transtemporal states} of a 
system described by a quantum history theory. \\ 

To get insight into the possible structure of 
general quantum 
history theories it is worthwhile to study 
the structure of the 
space of decoherence functionals for general 
quantum history 
theories in some detail. In particular it 
is of some interest 
to find operator representations for 
decoherence functionals. 
In the present work we consider the situation where 
that the space of  histories 
is given by the set of projection operators 
on some, in general, 
infinite dimensional Hilbert space. This 
choice can be motivated 
by appealing to the history formulation 
of standard quantum mechanics 
as given by Isham \cite{Isham94}. \\ 

Let $\mathcal{B}(\mathfrak{H})$ be the 
space of all bounded 
operators on a Hilbert space $\mathfrak{H}$ 
and let $\mathcal{P}(\mathfrak{H})$ be  
the lattice of projections in 
$\mathcal{B}(\mathfrak{H})$ (here
$\mathcal{P}(\mathfrak{H})$ is interpreted 
as the set of
histories). Then a {\emph{decoherence 
functional for}} 
$\mathfrak{H}$ is a complex valued 
function $d :
\mathcal{P}(\mathfrak{H}) \times 
\mathcal{P}(\mathfrak{H}) \to
\mathbb{C}$, defined on all
ordered pairs of projections in 
$\mathcal{P}(\mathfrak{H})$, such that
\begin{itemize} 
\item[(i)] \emph{Hermiticity:} 
$d(p,q) = d(q,p)^*$ 
for each $p$ and $q$ in 
$\mathcal{P}(\mathfrak{H})$. 
(Here $^*$ denotes complex conjugation.)
\item[(ii)] \emph{Positivity:} $d(p,p) 
\geq 0$ for each 
$p \in \mathcal{P}(\mathfrak{H})$. 
\item[(iii)] \emph{Normalisation:} $d(1,1) 
= 1.$  
\item[(iv)] \emph{Orthoadditivity:}
$d(p_1 + p_2, q) = d(p_1,q) + d(p_2, q)$ 
whenever $p_1 \in \mathcal{P}(\mathfrak{H})$ 
and $p_2 \in \mathcal{P}(\mathfrak{H})$ are 
perpendicular  
and $q \in \mathcal{P}(\mathfrak{H})$ is an 
arbitrary projection.
\end{itemize} 
There are stronger notions of orthoadditivity 
which are useful:
\begin{itemize} 
\item[(iv)$'$] \emph{Countable Additivity:} \\ 
A decoherence functional said to be 
\emph{countably additive} 
if, whenever 
$\{ p_i \}_{i \in \mathbb{N}}$ is a 
countable collection of pairwise 
orthogonal projections, for each $q \in 
\mathcal{P}(\mathfrak{H})$,
\begin{center} 
$d \left(\sum_{i=1}^{\infty} p_i,q \right) 
= \sum_{i=1}^{\infty}
 d(p_i,q)$. \end{center}
Here the series on the right hand side is 
rearrangement invariant 
and hence is absolutely convergent.
\item[(iv)$''$] \emph{Complete Additivity:} \\ 
A decoherence functional is said to be 
\emph{completely additive} 
if, whenever 
$\{ p_i \}_{i \in I}$ is an infinite 
collection of pairwise orthogonal 
projections, 
\begin{center} $d \left(\sum_{i \in I} p_i,q 
\right) = \sum_{i \in I} 
d(p_i,q),$ \end{center} for each 
$q \in \mathcal{P}(\mathfrak{H})$. 
Here the convergence is always absolute.
\end{itemize}    
When $\mathfrak{H}$ is finite dimensional 
and of dimension greater
than two, it follows from Isham, 
Linden and Schreckenberg \cite{IshamLS94} 
that for each  bounded decoherence 
functional $d$ on
$\mathcal{P}(\mathfrak{H}) \times 
\mathcal{P}(\mathfrak{H})$ there is 
a canonical operator $\mathfrak{X}$ 
on $\mathfrak{H} \otimes 
\mathfrak{H}$ such that, for all 
$p, q \in
\mathcal{P}(\mathfrak{H})$ 
\begin{equation} d(p,q) = 
\mathrm{tr}_{\mathfrak{H} \otimes \mathfrak{H}}
\left((p \otimes q) \mathfrak{X} \right). 
\label{E1} \end{equation} 
The properties (i), (ii) and (iii) of 
decoherence functionals imply: 
\begin{itemize} \item[(i)] 
$\mathrm{tr}_{\mathfrak{H} \otimes \mathfrak{H}}
\left((p \otimes q) \mathfrak{X} \right) 
= \mathrm{tr}_{\mathfrak{H} \otimes
\mathfrak{H}} \left((q \otimes p) 
\mathfrak{X}^{*} \right)$;
\item[(ii)] 
$\mathrm{tr}_{\mathfrak{H} \otimes 
\mathfrak{H}}\left((p 
\otimes p) \mathfrak{X} \right) \geq 0$;
\item[(iii)] 
$\mathrm{tr}_{\mathfrak{H} \otimes 
\mathfrak{H}}(\mathfrak{X}) =
1.$ \end{itemize}
Conversely, given any operator 
$\mathfrak{X}$ on the finite
dimensional space $\mathfrak{H} 
\otimes \mathfrak{H}$ which satisfies
(i), (ii), and (iii), there is a 
unique decoherence functional given by 
(\ref{E1}), see \cite{IshamLS94}. \\ 

Our aim here is to investigate to 
what extent the representation 
(\ref{E1}) remains valid when 
$\mathfrak{H}$ is infinite dimensional. 
If (\ref{E1}) is to hold for all $p$ 
and $q$ in 
$\mathcal{P}(\mathfrak{H})$, then $\mathfrak{X}$ 
must be a trace class 
operator. Moreover $d$ must be 
bounded and completely additive. 
 It would be pleasing if these 
 conditions were sufficient to 
imply the existence of a trace 
class operator which satisfies 
(\ref{E1}). Unfortunately this is 
false, since  there are physically 
natural examples of `well-behaved' 
decoherence functionals for 
which (\ref{E1}) fails. On the 
other hand we shall show that it is 
possible to generalise the representation 
(\ref{E1}) 
to infinite dimensions 
precisely when $d$ satisfies a 
sufficiently strong boundedness 
condition (\emph{tensor boundededness}). \\

Since many decoherence functionals of physical 
interest are not 
tensor bounded and hence not representable by 
trace class operators, 
we then identify the large and physically 
important class of 
decoherence functionals which can be represented, 
canonically, 
by bounded operators on $\mathfrak{H} \otimes 
\mathfrak{H}$.  
These are the \emph{tracially bounded} 
decoherence functionals.  We shall see that to 
each tracially 
bounded decoherence functional $d$ there 
corresponds a unique bounded 
linear operator $\mathfrak{M}$ on $\mathfrak{H} 
\otimes \mathfrak{H}$ 
such that,
whenever $p$ and $q$ are projections with 
finite dimensional range,
\begin{equation} 
d(p,q) = \mathrm{tr}_{\mathfrak{H} \otimes 
\mathfrak{H})}((p \otimes q) \mathfrak{M}). 
 \label{E6}
\end{equation}
In general, $\mathfrak{M}$ 
is not of trace class and the formula 
(\ref{E6}) has 
no meaning for projections of infinite 
dimensional range.  But, 
when $d$ is also countably additive and 
$\mathfrak{H}$ 
is separable, (\ref{E6}) 
can be used to calculate $d$. For, given 
$p$ and $q$ in 
$\mathcal{P}(\mathfrak{H})$, each 
of them can be  written as a countable sum 
of orthogonal  projections 
of finite rank, $p =  \sum_i p_i$ and 
$q = \sum_j q_j$ 
and we find that
\begin{equation} d(p,q) = 
d\left(\sum_i p_i, \sum_j q_j \right) = 
\sum\limits_{i=1}^{\infty } 
\sum\limits_{j=1}^{\infty } 
\mathrm{tr}_{\mathfrak{H} \otimes \mathfrak{H}} 
\left((p_i \otimes q_j) \mathfrak{M} \right).
\end{equation}
When $\mathfrak{H}$ 
is not separable analogous results 
hold if $d$ is completely 
additive. \\

Throughout this work we denote 
the inner product on a Hilbert space 
$\mathfrak{H}$ by 
$\langle \cdot, \cdot \rangle$ and we 
adopt the convention that the 
inner product $\langle \cdot, \cdot 
\rangle$ is linear in the first 
variable and conjugate linear in the 
second variable. 
\section{Bounded decoherence functionals}
A decoherence functional $d: 
\mathcal{P}(\mathfrak{H}) \times 
\mathcal{P}(\mathfrak{H}) \to \mathbb{C}$ 
is said  to be \emph{bounded} 
if the set of real numbers $\{ \vert d(p,q) 
\vert : p \in 
\mathcal{P}(\mathfrak{H}), q \in 
\mathcal{P}(\mathfrak{H})\}$ 
is bounded.

In the sequel we will make use of the 
following theorem which is a
special case of a more general result 
proved in Wright
\cite{Wright95}.
\begin{theo} Let $\mathfrak{H}$ 
be a Hilbert space which is either 
infinite 
dimensional or of finite dimension 
greater than two. Let 
$\mathcal{P}(\mathfrak{H})$ 
be the lattice of projections on 
$\mathfrak{H}$. 
Then a decoherence functional 
$d$ can be extended (uniquely) to a 
bounded bilinear form 
$\mathcal{D} : \mathcal{B}(\mathfrak{H}) 
\times
\mathcal{B}(\mathfrak{H}) \to \mathbb{C}$ if, 
and only if, $d$ is bounded. \label{T1} \end{theo}
By setting $\mathfrak{Q}(x,y) = 
\mathcal{D}(x,y^*)$ 
we can replace bilinear forms on 
$\mathcal{B}(\mathfrak{H}) \times 
\mathcal{B}(\mathfrak{H})$ 
by sesquilinear forms on 
$\mathcal{B}(\mathfrak{H})$. 
An immediate consequence 
of Theorem \ref{T1} is that bounded 
decoherence functionals for 
$\mathfrak{H}$ (where $\mathfrak{H}$ 
is not of dimension 2) 
are in one-to-one correspondence 
with those bounded Hermitian forms 
$\mathfrak{Q}$ on 
$\mathcal{B}(\mathfrak{H})$ for which 
$\mathfrak{Q}(1,1) 
= 1$ and $\mathfrak{Q}(p,p) \geq 0$ 
for all projections $p$; see
\cite{Wright95}.
\section{Bilinearity and linearity: 
Isham-Linden-Schreckenberg 
representability} 
Let $\mathfrak{H}$ be a Hilbert space and  
let $\mathcal{K}(\mathfrak{H})$ be the 
ideal of compact operators in 
$\mathcal{B}(\mathfrak{H})$. Then 
$\mathcal{K}(\mathfrak{H}) 
= \mathcal{B}(\mathfrak{H})$ if, 
and only if, $\mathfrak{H}$ is 
finite dimensional. \\ 

We shall need some basic facts on tensor 
products of operator 
algebras. For a particularly elegant 
account, from first principles, 
of tensor products of $C^*$-algebras 
see Wegge-Olsen
\cite{WeggeOlsen92} and, for 
a more advanced treatment, see 
Kadison and Ringrose
\cite{KadisonR8386}. \\

Let us recall that if $\mathfrak{H}$ 
is a Hilbert space, 
the algebraic tensor 
product $\mathfrak{H} \otimes_{alg} 
\mathfrak{H}$ 
can be equipped with an inner product 
such that 
$\langle h_1 \otimes f_1, h_2 \otimes f_2 
\rangle =  \langle h_1, 
h_2 \rangle \langle f_1, f_2 \rangle$. 
The completion 
of $\mathfrak{H} \otimes_{alg} \mathfrak{H}$ 
with respect to this inner product is 
the Hilbert space 
tensor product $\mathfrak{H} \otimes 
\mathfrak{H}$. 
When $x$ and $y$ are bounded operators 
on $\mathfrak{H}$, then 
there is a unique operator in 
$\mathcal{B}(\mathfrak{H} \otimes
\mathfrak{H})$, denoted by $x \otimes y$, 
such that 
$(x \otimes y)(h \otimes f) = x(h) 
\otimes y(f)$ for all $h$ and $f$
in $\mathfrak{H}$. \\

Let $\mathcal{A}$ and $\mathcal{B}$ 
be $C^*$-algebras of operators 
acting on $\mathfrak{H}$, then the 
\emph{algebraic tensor product} 
$\mathcal{A} \bullet \mathcal{B}$ 
can be identified with the 
$^*$ algebra, acting on 
$\mathfrak{H} \otimes \mathfrak{H}$, 
which consists of all finite sums of 
operators 
of the form $A \otimes B$, 
with $A \in \mathcal{A}$ and 
$B \in \mathcal{B}$. 
The norm closure of 
$\mathcal{A} \bullet \mathcal{B}$ 
is the \emph{$C^*$-tensor product of} 
$\mathcal{A}$ \emph{and} 
$\mathcal{B}$
and is denoted by $\mathcal{A} 
\otimes \mathcal{B}$. 
(This is 
also called the spatial $C^*$-tensor 
product  to distinguish it 
from other possible $C^*$-tensor products; 
see \cite{WeggeOlsen92}.) 
When $\mathcal{A}$ and 
$\mathcal{B}$ are von Neumann algebras of 
operators acting on 
$\mathfrak{H}$, then the 
closure of $\mathcal{A} \bullet \mathcal{B}$ 
in the weak operator topology of 
$\mathcal{B}(\mathfrak{H} \otimes
\mathfrak{H})$ is the \emph{von 
Neumann tensor product of} 
$\mathcal{A}$ \emph{and} $\mathcal{B}$, 
denoted 
by $\mathcal{A} \overline{\otimes} 
\mathcal{B}$. \\ 

The algebraic tensor product 
$\mathcal{K}(\mathfrak{H}) \bullet 
\mathcal{K}(\mathfrak{H})$ 
embeds naturally into 
$\mathcal{B}(\mathfrak{H}) \bullet 
\mathcal{B}(\mathfrak{H})$
which, in turn embeds naturally 
into $\mathcal{B}(\mathfrak{H}) 
\overline{\otimes} \mathcal{B}(\mathfrak{H}) 
= \mathcal{B}(\mathfrak{H}
\otimes \mathfrak{H})$ 
(see for instance Kadison and 
Ringrose \cite{KadisonR8386}, 
Chapter 11.2.). 
This embedding of 
$\mathcal{K}(\mathfrak{H}) \bullet 
\mathcal{K}(\mathfrak{H})$ in 
$\mathcal{B}(\mathfrak{H} \otimes
\mathfrak{H})$ induces a (unique) 
pre-$C^*$-norm 
on $\mathcal{K}(\mathfrak{H}) \bullet 
\mathcal{K}(\mathfrak{H})$. The 
(spatial) $C^*$-tensor product 
$\mathcal{K}(\mathfrak{H}) \otimes 
\mathcal{K}(\mathfrak{H})$ 
is the closure of 
$\mathcal{K}(\mathfrak{H}) \bullet 
\mathcal{K}(\mathfrak{H})$ in 
$\mathcal{B}(\mathfrak{H} \otimes
\mathfrak{H})$ with respect to 
this pre-$C^*$-norm 
and can be identified with 
$\mathcal{K}(\mathfrak{H} \otimes
\mathfrak{H})$. \\ 

Let $d$ be 
a bounded decoherence functional 
for $\mathfrak{H}$, where 
$\mathfrak{H}$ is not of 
dimension two. 
Then, by Theorem \ref{T1}, 
$d$ has a unique extension to a 
bounded bilinear form 
$\mathcal{D} : \mathcal{B}(\mathfrak{H}) 
\times
\mathcal{B}(\mathfrak{H}) \to 
\mathbb{C}$ such that $d(p, q) = 
\mathcal{D}(p, q)$ for all $p$ and 
$q$ in $\mathcal{P}(\mathfrak{H})$. 
\\ 

Let $\mathcal{D}_{\mathcal{K}}$ the 
restriction of $\mathcal{D}$ to 
$\mathcal{K}(\mathfrak{H}) \times 
\mathcal{K}(\mathfrak{H})$. 
Then, by the fundamental property 
of the algebraic tensor product there 
is a unique linear functional 
$\beta : \mathcal{K}(\mathfrak{H})
\bullet \mathcal{K}(\mathfrak{H}) 
\to \mathbb{C}$ such that 
\begin{equation} \beta(x \otimes y) = 
\mathcal{D}_{\mathcal{K}}(x,y) = 
\mathcal{D}(x,y), \label{E2} 
\end{equation} \nopagebreak 
for all $x, y \in 
\mathcal{K}(\mathfrak{H})$.
In particular $d(p, q) = 
\beta(p \otimes q)$ for all projections $p$
and $q$ in 
$\mathcal{K}(\mathfrak{H})$. \\ \\ 
\textbf{Definition} 
\emph{The decoherence functional 
$d$ is said to be \textsf{tensor 
bounded} if the associated functional 
$\beta$ is bounded on 
$\mathcal{K}(\mathfrak{H}) \bullet 
\mathcal{K}(\mathfrak{H})$, 
when $\mathcal{K}(\mathfrak{H}) 
\bullet \mathcal{K}(\mathfrak{H})$ 
is equipped with its unique 
pre-$C^*$-norm.} 
\begin{lem} Let $\mathfrak{H}$ 
be an arbitrary Hilbert space. 
Let $\phi$ be a bounded 
linear functional on 
$\mathcal{K}(\mathfrak{H})$. 
Then \emph{(1)} there exists a unique trace 
class operator $T$ in 
$\mathcal{B}(\mathfrak{H})$ such that, 
for each $z \in \mathcal{K}(\mathfrak{H})$, 
\[ \phi(z) = \mathrm{tr}_{\mathfrak{H}}(zT). \] 
Furthermore \emph{(2)}, 
there is a unique extension of $\phi$ 
to an ultraweakly 
continuous functional 
$\widetilde{\phi}$ on 
$\mathcal{B}(\mathfrak{H})$ 
such that, for each $z \in 
\mathcal{B}(\mathfrak{H})$,
\[ \widetilde{\phi}(z) = 
\mathrm{tr}_{\mathfrak{H}}(zT). 
\] \label{L9} \end{lem} 
\textbf{Proof}: For (1) see 
\cite{Takesaki79}, page 63, 
or \cite{Schatten70}, 
page 48, Theorem 3 and for (2) see 
\cite{KadisonR8386}, Vol. II, page 749. 
\hfill $\Box$ \\ 

\begin{theo} \label{T2} 
Let $\mathfrak{H}$ be a Hilbert space 
which is not of dimension 
two. Let $d$ be a bounded decoherence 
functional for $\mathfrak{H}$. 
Then 
$d$ is tensor bounded if, and only if, 
there exists a trace class 
operator $\mathfrak{X}$ on 
$\mathfrak{H} \otimes \mathfrak{H}$ such that
\begin{equation} d(p,q) = 
\mathrm{tr}_{\mathfrak{H} \otimes
\mathfrak{H}} ((p \otimes q) 
\mathfrak{X}) \label{E3} \end{equation} 
for all projections $p$ and $q$ 
in $\mathcal{K}(\mathfrak{H})$.
\end{theo}  
\textbf{Proof}: Let $\mathcal{D}$ 
be the bounded bilinear form 
corresponding to $d$ and let 
$\beta$ be the associated linear
functional on the algebraic tensor 
product $\mathcal{K}(\mathfrak{H})
\bullet \mathcal{K}(\mathfrak{H})$. 

Let $\mathfrak{X}$ be a trace class 
operator which 
implements (\ref{E3}). 
Let $\phi(z) = \mathrm{tr}_{\mathfrak{H} 
\otimes \mathfrak{H}}(z
\mathfrak{X})$ 
for $z \in 
\mathcal{K}(\mathfrak{H}) \bullet 
\mathcal{K}(\mathfrak{H})$. 
Then $\phi(x \otimes y) = \mathcal{D}(x,y)$ for 
all $x, y \in \mathcal{K}(\mathfrak{H})$. 
Since, by the fundamental property of the 
algebraic tensor product, $\beta$ 
is the unique functional with this 
property, it coincides with $\phi$. 
So $\beta$ is bounded. \\ 
Conversely, suppose that $d$ is 
tensor bounded.  Then, by definition, 
$\beta$ is bounded. So it has a 
unique extension to a bounded linear 
functional $\widetilde{\beta}$ on 
$\mathcal{K}(\mathfrak{H} \otimes 
\mathfrak{H})$, 
the norm closure of 
$\mathcal{K}(\mathfrak{H}) \bullet 
\mathcal{K}(\mathfrak{H})$. 
By Lemma \ref{L9}(1) 
there exists a trace class operator 
$\mathfrak{X}$ 
on $\mathfrak{H} \otimes
\mathfrak{H}$ such that 
\[ d(p,q) = \mathcal{D}(p,q) = 
\beta(p \otimes q) 
= \mathrm{tr}_{\mathfrak{H} 
\otimes \mathfrak{H}}((p \otimes q)
\mathfrak{X}), \] 
for all projections $p,q \in 
\mathcal{K}(\mathfrak{H})$. 
\hfill $\Box$ \\ 
\begin{co} Let $\mathfrak{H}$ be a 
Hilbert space which is not of
dimension two. 
Let $d$ be completely additive. There 
exists a trace 
class operator $\mathfrak{X}$ on 
$\mathfrak{H} \otimes \mathfrak{H}$ such that
\begin{equation} d(p,q) = 
\mathrm{tr}_{\mathfrak{H} \otimes
\mathfrak{H}} \left((p \otimes q) 
\mathfrak{X} \right), \label{E4} 
\end{equation}
for all projections $p,q$ in 
$\mathcal{B}(\mathfrak{H})$ 
if, and only if, $d$ is tensor 
bounded. \end{co} 
\textbf{Proof}: When there exists a 
trace class operator 
$\mathfrak{X}$ such that (\ref{E4}) 
holds, then, by Theorem 
\ref{T2}, $d$ is tensor bounded. \\ 
Conversely, when $d$ is tensor bounded, 
the existence of $\mathfrak{X}$ 
such 
that (\ref{E4}) holds for all 
projections of finite rank is guaranteed 
by Theorem \ref{T2}. By appealing 
to the complete additivity of $d$ and the 
ultraweak continuity of the map 
$z \mapsto \mathrm{tr}_{\mathfrak{H}
\otimes \mathfrak{H}}(z \mathfrak{X})$ 
it is straightforward 
to establish (\ref{E4}) for arbitrary 
projections. \hfill $\Box$ \\ 
\begin{co} 
Let $\mathfrak{H}$ be a Hilbert space 
which is not of dimension two.
There is a one-to-one correspondence 
between completely additive, 
tensor bounded decoherence functionals 
$d$ for $\mathfrak{H}$ 
and trace class operators $\mathfrak{X}$ 
on $\mathfrak{H} \otimes \mathfrak{H}$ 
according to the rule 
\begin{equation} d(p,q) = 
\mathrm{tr}_{\mathfrak{H} \otimes
\mathfrak{H}} \left((p \otimes q) \mathfrak{X} 
\right), \label{E13} \end{equation} 
for all projections $p,q \in 
\mathcal{K}(\mathfrak{H})$ 
with the restriction that 
\begin{itemize} \item 
$\mathrm{tr}_{\mathfrak{H} \otimes 
\mathfrak{H}} \left((p \otimes q) 
\mathfrak{X} \right) = 
\mathrm{tr}_{\mathfrak{H} \otimes 
\mathfrak{H}} \left((q \otimes p) 
\mathfrak{X}^* \right)$; 
\item $\mathrm{tr}_{\mathfrak{H} 
\otimes \mathfrak{H}}((p 
\otimes p) \mathfrak{X})
\geq 0$;
\item $\mathrm{tr}_{\mathfrak{H} \otimes 
\mathfrak{H}}(\mathfrak{X}) 
= 1$. \end{itemize} \end{co} 
\textbf{Proof}: Straightforward. 
\hfill $\Box$ \\ \\
\textbf{Remark} The 
Isham-Linden-Schreckenberg Theorem follows 
immediately 
since, when $\mathfrak{H}$ is finite 
dimensional, 
$\mathcal{K}(\mathfrak{H}) \bullet 
\mathcal{K}(\mathfrak{H}) 
= \mathcal{K}(\mathfrak{H}) \otimes 
\mathcal{K}(\mathfrak{H}) = 
\mathcal{B}(\mathfrak{H}) \overline{\otimes} 
\mathcal{B}(\mathfrak{H}) = 
\mathcal{B}(\mathfrak{H} \otimes
\mathfrak{H})$ 
which is finite dimensional and 
every linear functional on a 
finite dimensional normed space 
is bounded. \\

\noindent This note is concerned 
with how far 
the Isham-Linden-Schreckenberg 
Theorem can be extended to infinite 
dimensions. Let $d$ be a bounded 
decoherence functional for $\mathfrak{H}$ 
where $\mathfrak{H}$ has
dimension greater than two. 
Let us call $d$ 
\emph{Isham-Linden-Schreckenberg-representable} 
(or, more shortly, \emph{ILS-representable}) 
if there 
exists a trace class operator $\mathfrak{X} 
\in \mathcal{B}(\mathfrak{H}
\otimes \mathfrak{H})$ such that 
\[ d(p,q) = \beta(p \otimes q) = 
\mathrm{tr}_{\mathfrak{H} \otimes
\mathfrak{H}}((p \otimes q) \mathfrak{X}), \] 
for all projections $p,q \in 
\mathcal{P}(\mathfrak{H})$. 
It follows from the results 
given above that a completely additive 
decoherence 
functional $d$ is ILS-representable if, 
and only if, it 
is tensor bounded. \\ \\ 
It would be pleasing if all countably 
additive decoherence functionals 
for $\mathfrak{H}$ (for 
$\mathfrak{H}$ separable and infinite 
dimensional) 
were ILS-representable. This is very far 
from the truth. The 
following example shows that there are 
very natural, `well-behaved' 
decoherence functionals which are not 
ILS-representable. 
\begin{ex} \label{Ex1} 
Let $\mathfrak{H}$ be a separable, infinite 
dimensional Hilbert space. 
Let $\psi$ be a unit vector in $\mathfrak{H}$. 
Let $B_{\psi} : \mathcal{B}(\mathfrak{H}) 
\times \mathcal{B}(\mathfrak{H}) \to 
\mathbb{C}$ be defined by 
\[ B_{\psi}(x,y) := \langle x \psi,y^{*} 
\psi \rangle \] 
and let $d_\psi$ be the (bounded) 
countably additive decoherence functional 
obtained by restricting $B_{\psi}$ to 
pairs of projections in 
$\mathcal{B}(\mathfrak{H})$. Then 
$d_\psi$ is not ILS-representable. \end{ex} 
\textbf{Proof}: As remarked above, 
there is a unique linear functional 
$\beta_{\psi} : \mathcal{K}(\mathfrak{H}) 
\bullet \mathcal{K}(\mathfrak{H})
\to \mathbb{C}$ such that $B_{\psi}(x,y) 
= \beta_{\psi}(x \otimes y)$ 
for each $x$ and $y$. By Proposition 0 in 
\cite{Wright97} $\beta_{\psi}$ is not bounded. 
Thus $d$ is not tensor bounded and so $d$ is 
not ILS-representable. \hfill $\Box$ \\ \\ 
It was shown in \cite{Wright97} that, 
for each $S \in
\mathcal{K}(\mathfrak{H}) \bullet 
\mathcal{K}(\mathfrak{H}),$
\[ \beta_{\psi}(S) = \sum_{i=1}^{\infty} 
\left\langle S (\psi 
\otimes \psi_i), \psi_i \otimes \psi \right\rangle, \] 
where $\{ \psi_i \}_{i \in \mathbb{N}}$ is an orthonormal basis for
$\mathfrak{H}$ with $\psi_1 = \psi$. 
Let $U$ be 
the unitary on $\mathfrak{H} 
\otimes \mathfrak{H}$ 
which maps $\psi_i \otimes 
\psi_j$ to $\psi_j \otimes \psi_i$, 
for each $i,j$. Let $P$ be 
the projection on $\mathfrak{H} 
\otimes \mathfrak{H}$ 
whose range is spanned by $\{\psi 
\otimes \psi_i : i = 1,2... \}$. 
Then $\beta_{\psi}(S) = 
\sum\limits_{i=1}^{\infty} 
\sum\limits_{j=1}^{\infty} \langle 
SPU( \psi_i \otimes \psi_j), 
(\psi_i \otimes \psi_j) \rangle$. 
Hence, when $S$ is of trace 
class, $\beta_{\psi}(S) = 
\mathrm{tr}_{\mathfrak{H} \otimes
\mathfrak{H}}(SPU)$. So, see 
\cite{Takesaki79} p.~320, 
when $S$ is positive 
\[ \vert \beta_{\psi}(S) \vert 
\leq \Vert PU \Vert \, \,
\mathrm{tr}_{\mathfrak{H} \otimes \mathfrak{H}}(S). 
\] 
It follows that $\beta_{\psi}$ 
is bounded on the rank one 
projections in $\mathcal{K}(\mathfrak{H})
\bullet \mathcal{K}(\mathfrak{H})$. \\ \\
\textbf{Remark}: Although 
$\sum\limits_{i=1}^{\infty } 
\sum\limits_{j=1}^{\infty} \langle
PU(\psi_i \otimes \psi_j), 
(\psi_i \otimes \psi_j) \rangle$ 
converges and has the value 
one, $PU$ is not of trace class 
because $PU(PU)^* = P$, where 
$P$ is a projection of non-finite rank.

\section{Tracially bounded decoherence 
functionals} 
In this section we consider a much 
larger class of decoherence 
functionals than those which are 
ILS-representable. As before, 
$\mathfrak{H}$ 
is a Hilbert space of dimension 
(finite or infinite) greater 
than two. Let $d$ be a bounded 
decoherence functional defined on 
$\mathcal{P}(\mathfrak{H}) \times 
\mathcal{P}(\mathfrak{H})$.  
Let $\mathcal{D}: 
\mathcal{B}(\mathfrak{H}) \times
\mathcal{B}(\mathfrak{H}) \to \mathbb{C}$ 
be the unique bounded bilinear 
form which extends $d$. Let $\beta$ 
be the unique linear functional on 
$\mathcal{K}(\mathfrak{H}) \bullet 
\mathcal{K}(\mathfrak{H})$ 
such that 
\[ \beta(x \otimes y) = 
\mathcal{D}(x,y) \] 
for all $x$ and $y$ in 
$\mathcal{K}(\mathfrak{H})$. \\

Before defining tracial boundedness, we 
introduce some notation. 
Let $\mathfrak{V}$ 
be a Hilbert space and $\xi$ a unit vector 
in $\mathfrak{V}$.  
Then the 
(rank one) projection from $\mathfrak{V}$ 
onto the one dimensional subspace 
spanned by $\xi$, will be denoted by $p_{\xi}$.  
Thus $p_\xi( \eta) = \langle \eta, 
\xi \rangle \xi$ for each 
$\eta \in \mathfrak{V}$. 
We observe that if $\alpha, \gamma$ 
are in $\mathfrak{H}$,
then $p_{\alpha \otimes \gamma} = 
p_\alpha \otimes p_\gamma$. 
We may identify $\mathfrak{H} 
\otimes_{alg} \mathfrak{H}$ 
with the dense subspace of $\mathfrak{H} 
\otimes \mathfrak{H}$ 
consisting of those vectors 
which are finite sums of elementary tensors. 
When $\xi$ is in 
$\mathfrak{H} \otimes_{alg} \mathfrak{H}$, 
then $p_{\xi}$ is in 
$\mathcal{K}(\mathfrak{H}) \bullet
\mathcal{K}(\mathfrak{H})$. \\ \\ 
\textbf{Definition} 
\enlargethispage{0.7cm}
\emph{The decoherence functional 
$d$ is said to be 
\textsf{tracially bounded} 
if it is bounded and, when $\beta$ 
is the corresponding linear functional 
on $\mathcal{K}(\mathfrak{H}) \bullet 
\mathcal{K}(\mathfrak{H})$, 
there exists a constant $C$ such that, 
for each unit 
vector $\xi$ in $\mathfrak{H} \otimes_{alg} 
\mathfrak{H}$,
$\vert \beta(p_\xi) \vert \leq C$.} \\ 

It is clear that the decoherence functional 
$d_{\psi}$ 
considered in Example \ref{Ex1} 
is tracially bounded.  Hence tracial 
boundedness is a strictly 
weaker condition than tensor boundedness. \\ 

The following technical lemma must be 
well know but, since we 
do not know of a convenient reference, 
an argument is supplied 
for the convenience of the reader.
\begin{lem} Let $\mathcal{L}$ be a 
bounded operator on 
$\mathfrak{H} \otimes
\mathfrak{H}$ such that, for all 
$\alpha, \beta \in \mathfrak{H}$,
\[ \langle 
\mathcal{L}(\alpha \otimes \beta), 
\alpha \otimes \beta \rangle = 0. 
\] \label{L0} 
Then $\mathcal{L} = 0$. \end{lem} 
\textbf{Proof}: Let $\Phi(\alpha, 
\beta, \gamma, \delta) 
= \langle 
\mathcal{L}(\alpha \otimes \beta), \gamma 
\otimes \delta \rangle$. 
So $\Phi$ is linear in the first two 
variables and conjugate linear in the 
third and fourth variables 
and \\ \parbox{16cm}{
\[ \Phi(\alpha, \beta, \alpha, \beta) = 0 
\quad \mathrm{ for } \quad \! \! 
\mathrm{ all } \quad 
\alpha, \beta \in \mathfrak{H}. \]} 
\begin{minipage}{1cm} 
\hspace*{\fill} (i)
\end{minipage} \\ 
On replacing $\alpha$ in (i) by $\alpha + \alpha'$ 
and expanding, using 
the linearity in the first variable and 
the conjugate linearity 
in the third variable, we obtain, after 
applying the identity 
(i) to two of the terms, \\ \parbox{16cm}{
\[ \Phi(\alpha, \beta, \alpha', \beta) + 
\Phi(\alpha', \beta,
\alpha, \beta) = 0. \]} 
\begin{minipage}{1cm} \hspace*{\fill} 
(ii) \end{minipage} \\ 
On replacing $\alpha$ by $i \alpha$ 
in (ii) we obtain
\[ i \Phi(\alpha, \beta, \alpha', 
\beta) - i \Phi(\alpha', \beta,
\alpha, \beta) = 0. \]
So \\ \parbox{16cm}{\[ \Phi(\alpha, 
\beta, \alpha', \beta) = 0. \]} 
\begin{minipage}{1cm} \hspace*{\fill} 
(iii) \end{minipage} \\ 
We replace $\beta$ by $\beta + \beta'$ 
in (iii), 
expand, and apply (iii) to two of 
the terms, obtaining \\ \parbox{16cm}{
\[ \Phi(\alpha, \beta', \alpha', \beta) + 
\Phi(\alpha, \beta,
\alpha', \beta') = 0. \]} \begin{minipage}{1cm}
\hspace*{\fill} (iv) \end{minipage} \\ 
On replacing $\beta'$ by $i \beta'$ in (iv), 
dividing by i and subtracting the 
result from (iv) we obtain
\[ 0 = \Phi(\alpha, \beta, \alpha', 
\beta') = \langle 
\mathcal{L} (\alpha
\otimes \beta), \alpha' \otimes 
\beta' \rangle. \] 
Hence $ \langle \mathcal{L} 
\xi, \eta \rangle =0$ for all $\xi, 
\eta$ in 
$\mathfrak{H} \otimes_{alg} 
\mathfrak{H}$. Since $\mathcal{L}$ 
is bounded, this 
implies that $\mathcal{L} = 0$. 
\hfill $\Box$
\begin{prop} Let $\mathfrak{H}$ be 
a Hilbert space of 
dimension greater than two and let \linebreak[4]
\mbox{$d : \mathcal{P}(\mathfrak{H}) 
\times \mathcal{P}(\mathfrak{H}) \to
\mathbb{C}$} be a bounded decoherence 
functional for $\mathfrak{H}$. 
Then there exist families of trace 
class operators 
$\{ \mathfrak{X}_i \}_{i \in I}$ and 
$\{ \mathfrak{Y}_i \}_{i \in I}$ 
on $\mathfrak{H}$, where, for each 
$x$ and $y$ in 
$\mathcal{K}(\mathfrak{H})$, 
$\sum_{i \in I} \vert 
\mathrm{tr}_{\mathfrak{H}}(x 
\mathfrak{X}_i) \vert^2$ 
and $\sum_{i \in I} \vert 
\mathrm{tr}_{\mathfrak{H}}(y 
\mathfrak{Y}_i) \vert^2$ are convergent 
and, for all $S \in
\mathcal{K}(\mathfrak{H}) \bullet 
\mathcal{K}(\mathfrak{H})$,
\begin{equation} 
\beta(S) = \sum_{i \in I} 
\mathrm{tr}_{\mathfrak{H} \otimes \mathfrak{H}}
\left(S \left(\mathfrak{X}_i \otimes 
\mathfrak{X}^{*}_i - 
\mathfrak{Y}_i \otimes 
\mathfrak{Y}^{*}_i \right) \right), 
\end{equation} 
where the infinite series is absolutely 
convergent. 
\label{P2} \end{prop}
\textbf{Proof}: This is an easy 
consequence of Theorem 6 in
\cite{Wright97}. \hfill $\Box$ \\ \\ 
In the following we shall now 
suppose that $\mathfrak{H}$ 
is separable.  All 
our results can be extended to 
general Hilbert spaces but the 
notation becomes simpler and more 
transparent when $\mathfrak{H}$ 
is separable. \\

Let $d, \beta$ be as in Proposition 
\ref{P2} and 
let $\xi$ be a unit vector in 
$\mathfrak{H} \otimes_{alg}
\mathfrak{H}$. So $\xi$ is a finite 
sum of simple 
tensors $\alpha_i \otimes \beta_i$, 
where each $\alpha_i$ and 
$\beta_i$ is in $\mathfrak{H}$. Then 
Proposition \ref{P2} 
implies that
\begin{eqnarray} \nonumber \beta(p_\xi) & = & 
\sum\limits_{i=1}^{\infty} 
\mathrm{tr}_{\mathfrak{H} \otimes
\mathfrak{H}} \left(p_\xi 
\left(\mathfrak{X}_i \otimes
\mathfrak{X}_i^* - \mathfrak{Y}_i 
\otimes
\mathfrak{Y}_i^* \right) \right) \\
& = & \sum\limits_{i=1}^{\infty } 
\left\langle
\left( \mathfrak{X}_i \otimes 
\mathfrak{X}_i^* - 
\mathfrak{Y}_i \otimes \mathfrak{Y}_i^* 
\right) \xi, \xi
\right\rangle. \label{E7} \end{eqnarray} 
Let us now assume that $d$ is tracially 
bounded.  Then, using
(\ref{E7}) and polarisation, 
we find that there exists a constant $C$ such 
that \[ \left\vert \sum\limits_{i=1}^{\infty } 
\left\langle \left( \mathfrak{X}_i 
\otimes \mathfrak{X}_i^* - 
\mathfrak{Y}_i \otimes 
\mathfrak{Y}_i^* \right) \xi, \eta
\right\rangle \right\vert \leq C \, \, 
\vert \vert \xi \vert \vert \quad 
\! \! \! \vert
\vert \eta \vert \vert \] for all $\xi, 
\eta$ in $\mathfrak{H}
\otimes_{alg} \mathfrak{H}$. \\ 
It follows from this that there is a 
bounded linear operator 
$\mathfrak{M}$ on $\mathfrak{H} \otimes 
\mathfrak{H}$ such that
\[ \langle \mathfrak{M} \xi, \eta \rangle = 
\sum\limits_{i=1}^{\infty} 
\left\langle \left( 
\mathfrak{X}_i \otimes 
\mathfrak{X}_i^* - \mathfrak{Y}_i 
\otimes \mathfrak{Y}_i^* \right) 
\xi, \eta \right\rangle 
\] for all $\xi, \eta$ in 
$\mathfrak{H} \otimes_{alg} \mathfrak{H}$. 
In particular, for $\alpha$ 
and $\gamma$ unit vectors in
$\mathfrak{H}$, 
\[ \beta \left(p_{\alpha} 
\otimes p_{\gamma} \right) = 
\beta \left(p_{\alpha \otimes 
\gamma} \right) = 
\langle \mathfrak{M}(\alpha \otimes 
\gamma), \alpha \otimes \gamma
\rangle = \mathrm{tr}_{\mathfrak{H} 
\otimes
\mathfrak{H}}(\mathfrak{M} p_{\alpha 
\otimes \gamma}) = 
\mathrm{tr}_{\mathfrak{H} \otimes \mathfrak{H}}
(\mathfrak{M}(p_\alpha \otimes p_\gamma)). \] 
Hence, by orthoadditivity, when 
$p$ and $q$ are projections of finite 
rank on $\mathfrak{H}$ 
\[ d(p,q) = \beta(p \otimes q) = 
\mathrm{tr}_{\mathfrak{H} \otimes
\mathfrak{H}} \left(\mathfrak{M}(p 
\otimes q) \right). \] 
\begin{prop} Let the decoherence 
functional $d$ be tracially 
bounded. Then there exists a unique 
bounded linear operator 
$\mathfrak{M}$ on $\mathfrak{H} 
\otimes \mathfrak{H}$ such that
\begin{equation} d(p,q) = 
\mathrm{tr}_{\mathfrak{H} \otimes
\mathfrak{H}} \left( \mathfrak{M}(p 
\otimes q) \right) \label{E8} 
\end{equation} 
whenever $p$ and $q$ are finite rank 
projections on $\mathfrak{H}$.
\label{P3} \end{prop} 
\textbf{Proof}: 
The preceding argument establishes 
the existence of $\mathfrak{M}$ 
with the required properties. If 
$\widehat{\mathfrak{M}}$  also 
satisfies (\ref{E8}), then, for all 
$\alpha$ and $\beta$ in
$\mathfrak{H}$, 
\[ \left\langle \left( \mathfrak{M} - 
\widehat{\mathfrak{M}} \right)
(\alpha \otimes \beta), \alpha 
\otimes \beta \right\rangle = 
\mathrm{tr}_{\mathfrak{H} \otimes 
\mathfrak{H}} \left( \left(
\mathfrak{M} - \widehat{\mathfrak{M}} 
\right) 
p_{\alpha \otimes \beta} \right) = 0. \] 
Hence, by Lemma \ref{L0}, 
$\mathfrak{M} - \widehat{\mathfrak{M}} = 0$. 
\hfill $\Box$ \\ \\
Let us recall that each projection in 
$\mathcal{B}(\mathfrak{H})$ 
is the sum of an orthogonal 
family of rank one projections.
\begin{theo} Let $d$ be a countably 
additive, tracially bounded 
decoherence functional for 
$\mathfrak{H}$, where 
$\mathfrak{H}$ is separable and of 
dimension greater than two. Then 
there exists a unique bounded 
linear operator $\mathfrak{M}$ on 
$\mathfrak{H} \otimes \mathfrak{H}$
such that, whenever $p$ and $q$ 
are projections 
in $\mathcal{P}(\mathfrak{H})$ and 
$\{p_n \}_{n \in \mathbb{N}}$ 
and $\{q_n \}_{n \in \mathbb{N}}$ 
are, respectively, 
orthogonal families of finite rank 
projections with 
$p = \sum_{n \in \mathbb{N}} p_n$ and 
$q = \sum_{n \in \mathbb{N}} q_n$, then
\begin{equation} d(p,q) = 
\sum\limits_{i=1}^{\infty} 
\sum\limits_{j=1}^{\infty} 
\mathrm{tr}_{\mathfrak{H} 
\otimes \mathfrak{H}} \left((p_i 
\otimes q_j) \mathfrak{M} \right). 
\label{E0} 
\end{equation} \end{theo} 
\textbf{Proof}: 
The countable additivity of $d$ and 
Proposition \ref{P3} imply 
the existence of a unique bounded 
linear operator $\mathfrak{M}$ 
satisfying (\ref{E0}). \hfill $\Box$ \\

When $\mathfrak{H}$ is infinite dimensional, 
it follows from
\cite{Wright96} that when $\mathfrak{H}$ 
is separable every countably
additive decoherence functional on 
$\mathcal{P}(\mathfrak{H})$ is
bounded and, when $\mathfrak{H}$ is 
not separable, every completely
additive decoherence functional on 
$\mathcal{P}(\mathfrak{H})$ is
bounded. By contrast, 
unbounded (`countably additive') decoherence
functionals exist on 
$\mathcal{P}(\mathfrak{H})$ whenever
$\mathfrak{H}$ is of finite dimension 
greater than one \cite{Wright96}. \\

When $\mathfrak{H}$ is not separable, 
then, provided $d$ 
is completely additive, 
the obvious analogue of (\ref{E0}) holds.

\end{document}